\documentclass[reprint,superscriptaddress,amsmath,amssymb,prl]{revtex4-1}

\usepackage{graphicx}

\usepackage[varg]{txfonts}

\newcommand{\be}[1]{\begin{equation}\label{#1}}
\newcommand{\ee}{\end{equation}}
\newcommand{\SI}[2]{{#1}~{\textrm{#2}}}

\newcommand{\tmix}{\tau_\textrm{mix}}
\newcommand{\tagg}{\tau_\textrm{cls}}
\newcommand{\tpro}{\tau_\textrm{pro}}
\newcommand{\Rend}{R_p^\textrm{end}}
\newcommand{\Rmix}{R_\textrm{mix}}
\newcommand{\cs}{c_s^\mathrm{free}}

\newcommand{\cmp}{c_{\text{m}p}} 

\newcommand{\kT}{k_\mathrm{B} T}

\newcommand{\ud}{\mathrm{d}}

\newcommand{\figref}[1]{Figure~\ref{#1}}

\begin{document}

\title{Controlled Nanoparticle Formation by Diffusion Limited Coalescence}

\def\dsmres{DSM Research, PO Box 18, NL-6160 MD Geleen, The Netherlands}
\author{R. Stepanyan}
  \affiliation{
    \dsmres}%
\author{J. G. J. L. Lebouille}
  \affiliation{
    \dsmres}%
\author{J. J. M. Slot}
  \affiliation{
    \dsmres}%
  \affiliation{%
    Department of Mathematics and Computer Science, 
    Eindhoven University of Technology, PO Box 513, 5600 MB, 
    Eindhoven, The Netherlands}%
\author{R. Tuinier}
  \affiliation{
    \dsmres}%
  \affiliation{%
    Van 't Hoff Laboratory, Debye Institute, Utrecht University, Padualaan 8, 3584 CH, Utrecht, The Netherlands}%
\author{M. A. Cohen Stuart}
  \affiliation{%
    Laboratory of Physical Chemistry and Colloid Science, 
    Wageningen University, Dreijenplein 6, 6703 HB,
    Wageningen, The Netherlands}%

\begin{abstract}
Polymeric nanoparticles (NPs) have a great application potential 
in science and technology.
Their functionality strongly depends on their size. 
We present a theory for the size of NPs formed by precipitation of polymers into 
a bad solvent in the presence of a stabilizing surfactant. 
The analytical theory is based upon diffusion-limited coalescence kinetics of the polymers. 

Two relevant time scales, a mixing and a coalescence time, are identified and 
their ratio is shown to determine the final NP diameter. 
The size is found to scale in a universal manner and is predominantly sensitive to 
the mixing time and the polymer concentration if the surfactant concentration 
is sufficiently high. 
The model predictions are in good agreement with experimental data. 
Hence the theory provides a solid framework for tailoring 
nanoparticles with a priori determined size.
\end{abstract}

\maketitle

Polymeric nanoparticles (NP) are gaining increasing attention
because of their numerous applications in, for instance, physics, chemistry and medicine
\cite{KNL:NATMAT:2:408, *PKH:NATNAN:2:751, *KHS:ACR:44:1016}.
The NP size and size distribution are the key parameters often determining 
their functionality. Therefore, one of the main experimental challenges is to prepare NPs
with well controlled dimensions tuned for a particular application. Models of NP 
formation, allowing one to steer the NP preparation process in the right direction, would 
simplify the size control significantly.

A high level of control over particle size is required in, for example,
targeted delivery (e.g., oncology). Size influences the
circulating half life time and
is crucial for selective cellular uptake: NPs between 
50 and \SI{200}{nm} in size are desired in passive cancer tumor 
targeting as they are too 
large to harm healthy cells but small enough to penetrate into the 
diseased ones. In brain imaging, fluorescent dye loaded
particles of about \SI{100}{nm} with biocompatible polymer coatings
are used because they produce small, 
sharply defined injection sites and show no toxicity
\emph{in vivo} or \emph{in vitro}
 \cite{GAF:PR:21:1428,KBD:NAT:310:498,AGG:AFM:19:2009}.

\begin{figure}[b]
\centering
\includegraphics[scale=1]{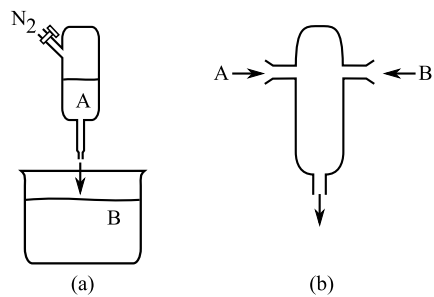}
\caption{Scheme of a pressure driven
injection device used in \cite{MGA:JPS:85:206} (a) 
and an impinging jets mixer used in \cite{JP:PRL:91:118302} (b). 
Fluid A is the organic phase comprising solvent, carrier polymer and 
the drug, 
fluid B is an aqueous solution of emulsifying agent.}
\label{fig:mixer}
\end{figure}

Although there are several methods for NP preparation, only few of them
permit high level of control on the particle size and the particle
size distribution \cite{GAF:PR:21:1428}. Often, a water insoluble 
moiety (e.g., a drug or a dye), needs to be encapsulated into a carrier 
polymer and protected by an emulsifying agent, which also makes the NP 
water soluble. 
In particular, the so-called nanoprecipitation method permits preparation
of nearly monodisperse NPs in a very simple and reproducible way \cite{NTH:JCR:25:89}. 
Typically, an organic phase, which is usually 
a dilute polymer solution, e.g., PCL in acetone, 
plus
the hydrophobic moiety 
to be encapsulated, e.g., a drug or a fluorescent dye, is injected by 
pressure into an aqueous solution of the emulsifying agent, 
\figref{fig:mixer}(a). 
As the organic solvent is chosen to be water-miscible, rapid quenching
(towards poor solvent conditions) 
of the hydrophobic polymer and the drug in water takes place. 
This results in 
coalescence
of the polymer and the drug into submicron particles decorated
by surfactant \cite{GAF:PR:21:1428,MGA:JPS:85:206}. 
Alternatively, a block copolymer can be used 
to combine the polymeric drug carrier and surfactant roles into a single component.
To promote a faster and better controlled mixing, 
an impinging jets mixer, \figref{fig:mixer}(b), has been employed 
in this case 
\cite{AGG:AFM:19:2009,JP:PRL:91:118302}. 

Despite the simplicity of the experimental method, one still lacks a 
comprehensive theoretical model that allows to predict 
how particle size depends on the materials and process
 parameters, in particular on concentration.
Therefore
in many practical situations 
investigators still
resort to simple empirical 
correlations \cite{S:CPS:273:505} or to statistical methods
such as experimental design \cite{MGA:JPS:85:206}. More advanced 
theoretical methods, such as Brownian dynamics simulations 
\cite{CHP:JCPB:112:16357} or population balance coupled to CFD simulation 
\cite{CF:IEC:49:10651},
do provide very valuable insights into the kinetics of mixing and 
rapid assembly upon quenching but do not permit formulation of a simple
yet physically meaningful analytical relationship between the 
experimentally relevant parameters and the NP size. 
Such a relationship would be extremely useful in designing 
NPs with \emph{a priori} determined size
as it would allow one to avoid a very laborious trial and error 
process.

In this Letter we formulate a model of the nanoprecipitation process.
The model grasps all essential features of the process and, at the same time,
provides a simple analytical expression for the NP size as a function 
of the mixing intensity and the surfactant and polymer properties.

We restrict ourselves to a bi-component system:
a dilute solution of a hydrophobic polymer is injected into a water/surfactant 
solution. 
Because the solvent and water are chosen to be well miscible, a rapid quench 
of the hydrophobic polymer in water takes place and the polymer particles 
start to coalesce upon encounter to 
form larger particles.
In parallel, the surfactant molecules, also subject to Brownian motion, 
adsorb on the surface on the newly formed polymeric particles 
and make their coalescence progressively
more difficult until a stable situation is reached.
As the NPs formed represent a system in a kinetically frozen state, 
their parameters will depend strongly on the  system kinetics, 
which includes at least three processes, namely 
(i) mixing of the polymer plus solvent with the aqueous surfactant solution 
taking place on the time scale $\tmix$; 
(ii) coalescence of the hydrophobic polymer particles in a hostile water environment, 
characterized by a time $\tagg$; 
and (iii) protection of the polymeric NPs by the surfactant taking 
place on the time scale $\tpro$ and bringing the system into a kinetically
frozen state.

Let us first consider the limit of `very fast mixing' $\tmix \to 0$, in the
absence of surfactant. The system then initially consists of collapsed polymer 
molecules homogeneously distributed in water. These molecules will diffuse, 
collide, and stick. If they would be hard particles, this would lead to fractal
aggregates, for which well-known growth laws have been developed. This case is
commonly known as `diffusion limited aggregation' (DLA). However, as our particles 
are liquid like, they will coalesce to homogeneous spherical particles rather than forming 
fractal aggregates, so that we deal 
with `diffusion limited coalescence' (DLC) \cite{A:PRL:81:4756}.
For this we have Smoluchowski theory \cite{S:ZPC:92:129,*DMP:PRL:105:120601}
with a rate $K=4\pi D^\prime R^\prime$, where $D^\prime$ and $R^\prime$ are 
the sum of the diffusion coefficients and the radii of the reacting species,
respectively.
Hence, in a mean field approximation, the polymer particle 
concentration $c_p$ is governed by a simple equation
\be{eq:1}
\frac{\ud c_p}{\ud t} 
  = -\frac{8}{3}\frac{\kT}{\eta} \, h \, c_p^2,
\ee
where the Stokes-Einstein expression 
$D_p=\kT / (6\pi\eta R_p)$
for the diffusion coefficient of a particle 
in a fluid with viscosity $\eta$ has been used. 
The factor $h$ accounts for the probability that a collision leads 
to a coalescence event.
 
The surfactant adsorbed on the particle surface
influences $h$, as it reduces the probability of a coalescence event 
to occur. Hence, $h$ is a function of the 
fraction of the particle surface protected by the surfactant,
$h \equiv h\left( {n(t) a^2}/{(4\pi R_p^2(t))} \right)$, where 
$n(t)$ denotes the average
number of surfactant molecules adsorbed on a polymer particle
with radius $R_p$
at time $t$, each surfactant molecule covering a surface area $a^2$.

Since we are dealing with coalescence rather than aggregation, 
there is a direct relation between particle mass and particle radius $R_p$
leading to the mass conservation law in the form
$c_p(t) R_p^3(t) = c_{p0} R_{p0}^3 $. 
Here, $R_{p0}$ and $c_{p0}$ are the size and the number 
concentration of the 
polymer particles immediately after the mixing took place.

In the absence of surfactant $h(\cdot) \equiv 1$ and \eqref{eq:1} can be recast 
in terms of the particle size yielding
$R_p^3(t) = R_{p0}^3 \left( 1+  t/\tagg \right)$,
with an encounter and coalescence time 
\be{eq:5}
\tagg = \frac{3}{8 c_{p0}} 
   \frac{\eta}{\kT}.
\ee

Assuming strong favorable interaction between polymer and surfactant,
the surfactant-polymer coagulation can be treated in a similar manner
\be{eq:8}
\frac{\ud \cs}{\ud t} = 
  -\frac{2}{3} \frac{\kT}{\eta} \left( \frac{1}{R_p}+\frac{1}{R_s} \right)
                                \left( R_p+R_s \right) \cs c_p h_s,
\ee
where $R_s$ and $\cs$ are the diffusion radius and the concentration of
the free (not adsorbed) surfactant molecules; $h_s=h_s(na^2/(4\pi R_p^2))$ denotes
the probability of adsorption. 
Equation \eqref{eq:8} is a straightforward generalization of \eqref{eq:1}, where
a relative diffusivity has been introduced as a sum of the polymer particles 
and the surfactant molecules diffusivities 
(see \cite{K:JCP:86:5052, *K:CR:87:167} for more details). 
Also, the reaction radius is assumed to equal $R_p+R_s$. 

An interesting observation at this point
is that $\tpro\sim\tagg$ and, hence, 
collision rate
of the polymer particles and their protection
by the surfactant go at approximately the same pace. Note, that we have neglected 
surfactant micellization by assuming that the surfactant molecules bound in surfactant micelles
behave similarly to the dissolved ones, at least in what concerns their agglomeration 
with polymeric NPs.

To describe the kinetics of coagulation, the exact functional form of $h$ and $h_s$ must be
specified and the equations \eqref{eq:1} and \eqref{eq:8}
need to be solved together. 
In fact, as we are only interested in the final particle 
size and not its time dependence, 
we can divide \eqref{eq:1} by \eqref{eq:8} yielding a single differential equation 
for $c_p$ as a function of $c_{s}^\text{free}$. Computing the exact form of 
$h(\cdot)$ can be quite involved,
although it is clear that $h(0)\simeq1$ and $h(1)\simeq 0$. The same holds for $h_s$. 
To simplify the matter significantly, we assume the 
surfactant adsorption not to influence the coalescence of particles until the 
particles are saturated with the surfactant and the coalescence is
stopped completely
\cite{LKS:PRL:98:036102}
and take $h(0\le x < 1)=1$ and $h(x\ge1)=0$ and the same for $h_s$. 
Such a choice  does
not change the scaling of all the important quantities but
implies that coagulation process stops when $n=4\pi R_p^2/a^2$ and, hence,
$c_{s\text{,end}}^\text{free} = c_{s0} - 4\pi (R_p^\text{end})^2 c_p^\text{end}/a^2$.
Solving the differential equation for $c_p$ as a function of $c_{s}^\text{free}$ 
explicitly and making 
use of the above relation between the end values of the concentrations and the radius, 
one derives a 
transcendent equation for the ratio $\zeta=\Rend/R_{p0}$ between the final and the
initial particle size
\be{eq:20}
1-\exp\left\{ 
  -\frac{3}{4}
  \left[ 
     \ln\zeta +\alpha(\zeta-1) + 
       \frac{1}{\alpha}\left( 1-\frac{1}{\zeta} \right)
  \right] \right\}
  = \frac{\kappa}{\zeta},
\ee
where $\alpha=R_{p0}/R_s$ is the ratio between the initial polymer particle
size and the diffusion radius of the surfactant and 
$\kappa= 4\pi R_{p0}^2 c_{p0}/(a^2 c_{s0})$
is the ratio of the total initial surface area on the polymer 
particles to the maximum area surfactant molecules can occupy and block.
   
Analytical solutions of \eqref{eq:20} are found for the limiting cases of
an excess of surfactant, $\kappa\ll 1$, and if surfactant is scarce,
$\kappa\gg 1$:   
\be{eq:21}
\Rend = R_{p0}\times\left\{ 
\begin{array}{ll} \displaystyle
 1+\frac{\kappa}{3/4 + \alpha + \alpha^{-1}} & \text{ if } \kappa \ll 1, \\
 \kappa                                      & \text{ if } \kappa \gg 1 .
\end{array}
\right.
\ee
This leads to a simple interpolation $\Rend \simeq R_{p0} (1+\kappa)$, which is surprisingly 
close to the exact numerical result.

A very peculiar implication of the fact that $\tagg\sim\tpro$, as pointed above, is that the final 
NP size does not depend on the mobility of polymer or surfactant molecules. The only
dominating factor in the `fast mixing' limit, when $\tmix\ll\tagg$, is the surfactant concentration.

Let us now consider the other limit, $\tmix\ge\tagg$, which is apparently characterized by a very
fast particle aggregation on the time scale shorter than the typical mixing time followed by
stabilization of the NPs' size at the times $t\ge\tmix$. 
Indeed, at the very beginning of the process, the polymers are present as isolated chains in a good solvent. 
As the solvent quality drops the polymers instantaneously collapse. Subsequent collision of collapsed chains 
leads to coalescence following the kinetics prescribed by \eqref{eq:1} with $h\equiv 1$.
Hence, the particle size at the end of the mixing, $t\simeq\tmix$, reads
$\Rmix \simeq R_{p0} \left( 1+{\tmix}/{\tagg}\right)^{1/3}$.
At longer times, 
there is enough time for the surfactant to adsorb onto the surface
of the coalescing polymer-rich dropets. Then
the system finds itself in a well mixed state and its 
kinetics obeys the set of equations
\eqref{eq:1} and \eqref{eq:8} as discussed above, but $\Rmix$ must be used as the `initial' particle 
size in \eqref{eq:20}.
This two-step process leads to a final expression for the polymer particle radius in a kinetically 
frozen state
\be{eq:30}
\Rend = R_{p0} (1+\kappa) \left( 1+\frac{\tmix}{\tagg}\right)^{1/3}.
\ee
It is characterized by a plateau at small $\tmix/\tagg$,
where the NP diameter is independent of mixing or encounter and coalescence time and is
totally governed by the surfactant concentration (parameter $\kappa$)
with the smallest particles obtained in excess of surfactant. 
The other regime, $\tmix/\tagg>1$, shows
a typical $1/3$ power law behavior and is dominated by the mixing efficiency.

Based on \eqref{eq:30} 
it follows
that for typical experimental conditions, 
i.e. excess of surfactant and relatively slow mixing, 
the final NP size depends mainly on the mixing time and 
the initial polymer mass concentration $\cmp$, 
$\Rend \propto (\cmp \tmix)^{1/3}$, 
and is independent of the polymer molar mass. 
Only a minor dependence on the molar mass of emulsifying agent can be observed
indirectly via $\tmix$, which can be sensitive to the viscosity of the surfactant solution. 
The same holds  for the temperature.

\begin{figure}[!t]
\centering
\includegraphics[scale=0.37]{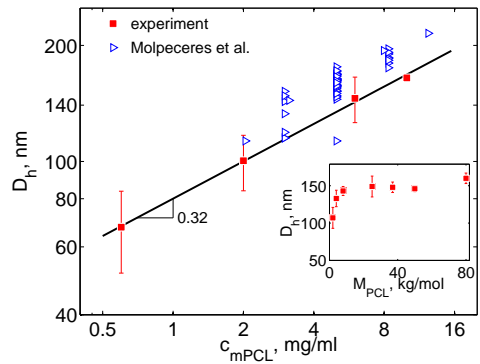}
\caption{Size of PCL (molar mass $M_w=\SI{25}{kg/mol}$) 
NPs prepared from an acetone solution quenched in an aqueous 
Pluronic solution
as
a function of the initial polymer concentration.
Solid line shows a fit to the data. 
Also the data from 
\citet{MGA:JPS:85:206} are shown. 
 Inset: Size of NP vs molar mass of PCL, at concentration  $c_{\text{m}PCL}=\SI{5}{mg/ml}$.}
\label{fig:exp:our}
\end{figure}

To appreciate the formula \eqref{eq:30} we compare its scaling predictions to our own experiments
as well as to the data available in the literature. 
In the nanoprecipitation experiments performed 
in our lab, 
PCL (CAPA 6250 supplied by Solvay) 
has been used as a carrier polymer and 
Pluronic 
(PF127
supplied by BASF) 
as a surfactant. 
PCL solution in acetone were quenched in a \SI{1}{wt\%} PF127 aqueous solution with a device 
similar to the one depicted in \figref{fig:mixer}(a). 
The hydrodynamic particle diameter $D_h$ has been measured by dynamic
light scattering.

As can be seen from \figref{fig:exp:our}, our results compare favorably 
to the data available in the literature
\cite{MGA:JPS:85:206} 
for the same system. As the experiments
are performed 
in the $\tmix>\tagg$ regime, the scaling obeys the $1/3$ power law 
as expected. To check the molar mass sensitivity, additional experiments have been performed
where PCL molar mass has been varied between 2 and \SI{80}{kg/mol}.
The diameter was, however, hardly affected by the molar mass
[see inset in \figref{fig:exp:our}],
in accord with the theoretical predictions.

\begin{figure}
\centering
\includegraphics[scale=0.5]{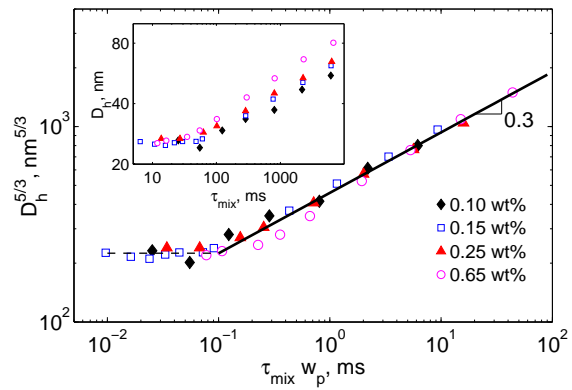}
\caption{Master curve of the size of the core of the diblock copolymer NPs \emph{versus} the 
rescaled mixing time, following the scaling predicted by \eqref{eq:30}. 
Inset: The original data from \citet{JP:PRL:91:118302}.}
\label{fig:exp:prudhomme}
\end{figure}

The data presented only cover the $\tmix > \tagg$ regime and
neither reach a particle size saturation limit at the very fast mixing, 
$\tmix < \tagg$, nor a crossover at $\tmix \sim \tagg$. However a very vast
experimental data set  is available for a somewhat
different system comprising a methanol solution of an amphiphilic diblock
copolymer (polybutylacrylate-\textit{b}-polyacrylic acid, each block \SI{7.5}{kg/mol}) 
quenched in water. 
By using a highly efficient impinging jet mixer, \figref{fig:mixer}(b),
 \citet{JP:PRL:91:118302} succeeded in covering a
very broad range of mixing times and observed all the three above-mentioned
regimes. Their original data -- the hydrodynamic diameters of the micelles formed
vs the mixing time -- are shown in the inset of \figref{fig:exp:prudhomme}. 
The coagulation in a dispersion containing diblock copolymers must obey kinetics very similar to the one
described by \eqref{eq:1} and, thus, yield scaling \eqref{eq:30} for
the NP size. 
This implies that a master curve
must be obtained in \figref{fig:exp:prudhomme}
if one shifts the data along the horizontal axis by the polymer 
mass fraction $w_p$. Moreover, a typical diameter scaling $(\tmix w_p)^{1/3}$
is expected to be observed at long mixing times.

One important difference between the concentration dependence of the size predicted in this Letter and the measurements in \cite{JP:PRL:91:118302}
is the fact that our equation \eqref{eq:30} does not take into account 
the size of the surfactant layer on top of a NP. 
Indeed, such an approximation certainly holds in case of a polymeric 
surfactant.
In case of diblock
copolymers, however, the size of the hydrophilic corona surrounding the 
hydrophobic core cannot be neglected.
To compute a hydrophobic core diameter from a hydrodynamic diameter of a 
copolymer micelle, we recall that the latter scales as a power 1/5 of the 
micelle mass \cite{DC:JP:43:531}. As the core of a micelle consists almost solely
of the hydrophobic polymer segments, the core size scales as a power 1/3 of the 
mass, yielding $R_\text{core} \propto D_h^{5/3}$.
The data redrawn in $D_h^{5/3}$ vs $\tmix c_p$ coordinates,
\figref{fig:exp:prudhomme}, indeed shows a master curve obeying equation
\eqref{eq:30}: it is characterized by a typical  $(\tmix/\tagg)^{1/3}$ scaling at long mixing
times and shows a plateau in the fast mixing regime, exactly as
the theory predicts.

Note, that the NP size in \figref{fig:exp:prudhomme} is completely determined by 
the kinetics and is not related to the equilibrium diblock copolymer micelle size. 
Indeed the latter would depend solely on the molar mass, composition and 
solvent quality, whereas the NP size is a strong function of concentration. 
Although the NP system is not in a thermodynamic equilibrium, it is long-lived. 
As an X-ray study on a somewhat different diblock copolymer system shows
\cite{LWM:PRL:102:188301}, micellization of copolymers includes two stages. The first 
rapid stage is totally controlled by kinetics and leads to NP formation described 
in the present work.
The second, several orders of magnitude slower process, drives the NP system to 
the thermodynamic equilibrium.

Based on the experimental evidence discussed above, one can conclude that
the nanoprecipitation model 
based on a diffusion limited coalescence mechanism
adequately describes the NP formation process. 
Two relevant time scales, the mixing and the encounter and coalescence times, 
are identified in \eqref{eq:30} and their ratio is
shown to be of a critical importance for the NP final diameter. 
The latter is predicted to scale in a universal manner and be sensitive
predominantly to the mixing time and the polymer concentration if the 
surfactant concentration is sufficiently high. 
The molar mass of the carrier polymer is shown to have little influence.
Experimental data available  corroborate the predictions of our model and 
provide a solid framework  for tailoring NPs with \emph{a priori}
determined size thus avoiding a laborious trial and error approach.

\acknowledgments
We thank Prof. R. Prud'homme for useful discussions.

%

\end{document}